\documentclass[
twocolumn,reprint,showpacs,preprintnumbers,aps,prl,letterpaper,superscriptaddress,nofootinbib,tightenlines,floats,floatfix]{revtex4-1}
\usepackage{graphicx}
\usepackage{dcolumn}
\usepackage{bm}
\usepackage{latexsym}
\usepackage{amsmath,amssymb}
\usepackage[draft=false]{hyperref}
\usepackage[latin1]{inputenc}
\usepackage{latexsym}
\usepackage{amsmath}
\usepackage{ifpdf}
\usepackage{color}
\usepackage{epstopdf}
\usepackage{subcaption}
\usepackage[skip=2pt,font=scriptsize]{caption}

\begin{document}
\title{Wormhole Cosmic Censorship: An Analytical Proof }

\author{Juan Carlos Del \'Aguila}
\email{jdelaguila@fis.cinvestav.mx}
\affiliation{Departamento de F\'isica, Centro de Investigaci\'on y de
  Estudios Avanzados del IPN, A.P. 14-740, 07000 M\'exico D.F., M\'exico.}
\author{Tonatiuh Matos}
\email{tmatos@fis.cinvestav.mx}
 \altaffiliation{Part of the Instituto Avanzado de Cosmolog\'ia (IAC)
  collaboration http://www.iac.edu.mx/}
\affiliation{Departamento de F\'isica, Centro de Investigaci\'on y de
  Estudios Avanzados del IPN, A.P. 14-740, 07000 M\'exico D.F.,
  M\'exico.}

\begin{abstract}
In this work we present an analytical proof of cosmic censorship in a Kerr-like phantom wormhole (WH) which contains a singularity that is not protected by an event horizon. We show that the naked singularity of this space-time is causally disconnected from the universe. To do so, we consider a slowly rotating limit and by means of the Hamilton-Jacobi theory separate the Hamiltonian of the geodesics into two polynomials. During this process we find a fourth conserved quantity. After examining the properties of these polynomials we conclude that the ring singularity is untouchable by any observer traveling in a geodesic of this space-time. We also derive the conditions on the four constants of motion that are necessary for a traveler to go back and forth both universes connected through the WH, and then compare its structure to that of a negative mass Kerr black hole.
\end{abstract}

\date{Received: date / Accepted: date}

\maketitle

\date{\today}


\maketitle

\section{I. Introduction}
In 1969, Roger Penrose studied gravitational collapse and suggested the existence of a "cosmic censor" which forbids the appearance of a naked singularity and hides it behind an event horizon \cite{penrose}. Many years later (1999), he discussed the question of cosmic censorship and provided some arguments against and in favor of it \cite{penrose2}, leading to the conclusion that the conjecture still did not possess a definite proof. In fact, the possibility that new mathematical techniques (twistor theory) were needed was also stated. Though there have been attempts to prove or disprove it, so far the question remains open. To this day, naked singularities persist as one of the biggest mysteries of General Relativity and little is known about the geometrical nature of this region of a space-time. Not to mention the physical unknowns it hides, since Einstein's field equations are not valid at the singularity and there is no physical theory that can describe it.

Even though the censorship mechanism is of general acceptance for black holes, since their singularities lie behind an event horizon, it remains to be seen if nature also conceals in some sort of way a naked singularity. Over the years, new types of solutions to Einstein's field equations have been found different from general black holes. One of these types of solutions are those which describe wormholes (WHs) connecting two different universes, to name a few examples \cite{ellis, morris}. These type of space-times become relevant to the cosmic censorship conjecture because, in most cases, are regular everywhere in space but contain naked singularities. In this paper we wonder if something just as cosmic censorship is present in a particular WH space-time.

In \cite{TM}, and later on in \cite{phantomWH}, the metric of a Kerr-like wormhole was reported as an exact solution of Einstein's field equations with a phantom like source. It is described by three parameters: mass, angular momentum and a scalar field charge. It also contains a ring singularity without an horizon. The line element in Boyer-Lindquist coordinates is given by

\begin{eqnarray}
ds^2=&-&f(dt+\Omega d\phi)^2  \nonumber\\
&+&\frac{1}{f}\left(\Delta\left(\frac{dl^2}{\Delta_1}+d\theta^2\right)+\Delta_1 \sin^2 \theta d\phi^2\right),
\label{ds2} 
\end{eqnarray}

where

\begin{eqnarray}
\Delta&=&(l-l_1)^2+(l^2_0-l^2_1)\cos^2\theta,
\label{delta}\\
\Delta_1&=&(l-l_1)^2+(l^2_0-l^2_1).
\label{delta1}
\end{eqnarray}


With line element components $\Omega$ and $f$

\begin{eqnarray}
\Omega=a\frac{(l-l_1)}{\Delta}\sin^2\theta, \hspace{0.6cm} f=\frac{(a^2+k^2_1)e^\lambda}{a^2+k^2_1e^{2\lambda}},
\label{omegaf}
\end{eqnarray}

where

\begin{equation}
\lambda=\frac{a^2+k^2_1}{2k_1\Delta}\cos\theta.
\label{lambda}
\end{equation}

The parameters $l_1$ and $l_0$ have units of distance ($l^2_0>l^2_1$) and are related to the size of the throat, while $a$ and $k_1$ are parameters with units of angular momentum. In Boyer-Lindquist coordinates, the ring singularity is located at $l=l_1$ and $\theta=\pi/2$. Alternatively, in the radial coordinate defined by $r^2=(l-l_1)^2+(l^2_0-l^2_1)$, it is located in the radius $r=\pm\sqrt{l^2_0-l^2_1}$. Also, the ADM mass and angular momentum were determined as $M=-l_1$ and $J=a$, respectively. 

\begin{figure}[htp]
	\centering
		\includegraphics[scale=0.43]{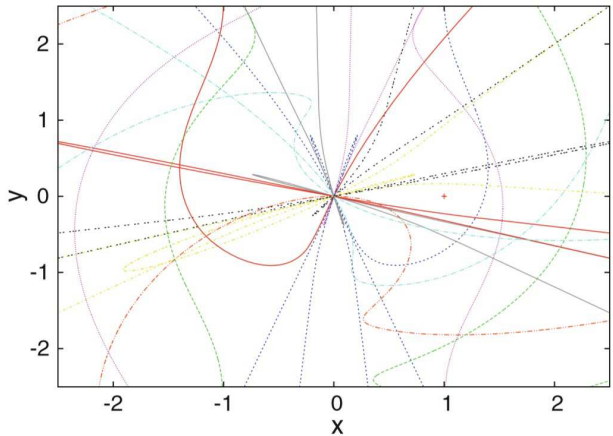} 
		\caption{Path of geodesics falling into the wormhole for different $\theta$, with: $l_1=1$, $l_0=1.1$, $a=0.1$, $k_1=0.11$, with the conserved quantities $\mathcal{E}=1$ and $\mathcal{L}=0.5$. The singularity is marked by a cross \cite{phantomWH}.}
	\label{fig:geo}
\end{figure}

Analysis of metric (\ref{ds2}) revealed that an observer could travel from one asymptotically flat universe to another through the WH and without facing strong tidal forces \cite{phantomWH}. Studying its geodesics numerically it was found that, as along as the observer does not travel on the equatorial plane, he could reach the throat of the WH. Fig.(\ref{fig:geo}) shows that none of the geodesic paths touch the ring singularity. In \cite{cosmic} it was conjectured that it could be possible to protect a singularity if it was surrounded with a WH's throat. However, this conclusion was obtained through the solution of the geodesic equations, which were solved by numerical methods. 

In this work we give an analytical proof, in the slowly rotating limit, that is not possible for a traveler to touch the ring singularity of this space-time, showing that the numerical results and the conjecture are valid. To do so, we perform a coordinate change in (\ref{ds2}) and, after deriving a condition or limit on some parameters of the space-time for a slowly rotating WH, apply said limit to the line element components (\ref{delta})-(\ref{lambda}). This will allow the Hamiltonian of the geodesics to be separable in the new coordinates system, thus obtaining a fourth conserved quantity through the Hamilton-Jacobi formalism. We also show an alternative path to finding this constant of motion through the existence of a second-rank Killing tensor. Afterward, we analyze the separated equations of motion and their relationship with the four conversed quantities, giving us insight on whether the singularity is touched by the observer or not. Finally, we briefly compare the structure of this WH space-time with a familiar example that also has a naked singularity, namely the negative mass Kerr black hole. Through this comparison we also wonder about the possible formation process of this WH.

Of course, what we present here is not a definite proof of the cosmic censorship conjecture and lacks the generality to do so. Nonetheless, the results obtained here are an example of a space-time with a  naked singularity whose geodesics are unable to touch it, i.e: the singularity is causally disconnected. This hints towards a different and particular case of the cosmic censorship hypothesis and could be considered as an argument in favor of it.

\section{II. Change of Coordinates and Slowly Rotating Limit}\label{sec:line}
Making $Lx=l-l_1$ and $y=\cos\theta$, with $L^2=l^2_0-l^2_1$, changes the line element (\ref{ds2}) and line element parameters (\ref{delta})-(\ref{lambda}) to

\begin{eqnarray}
ds^2=&-&f(dt+\Omega d\phi)^2 \label{ds2xy}\\
&+&\frac{1}{f}\left(\Delta\left(\frac{L^2dx^2}{\Delta_1}+\frac{dy^2}{1-y^2}\right)+\Delta_1(1-y^2)d\phi^2\right). \nonumber
\end{eqnarray}

Where now $\Delta=L^2(x^2+y^2)$, $\Delta_1=L^2(x^2+1)$, and

\begin{equation}
\Omega=\frac{ax(1-y^2)}{L(x^2+y^2)}, \hspace{0.6cm} \lambda=\frac{(a^2+k^2_1)y}{2k_1L^2(x^2+y^2)}.
\label{omegalambda}
\end{equation}

Notice that the throat of the WH now lies at $x=0$, and the ring singularity at $x=0$ and $y=0$. Also it is important to mention that $y=1$ is a coordinate singularity rather than a curvature singularity.

In order to determine the angular velocity of this space-time we consider a freely falling test particle in the WH. Its angular velocity $\omega$ can be obtained from 

\begin{equation}
\omega=\frac{\dot \phi}{\dot t}=\frac{g^{\phi \phi}p_\phi+g^{\phi t}p_t}{g^{t \phi}p_\phi+g^{tt}p_t},
\label{rot}
\end{equation}

where $p_\mu$ are the conjugate momenta of the test particle and $\dot{x^\mu}$ its coordinate velocities. As the particle moves across this space-time, its angular velocity will have contributions from both the rotating geometry of the WH and its own angular momentum, as can be seen from equation (\ref{rot}). Setting $p_\phi=0$ in (\ref{rot}) describes the angular velocity of a zero angular momentum particle. In this case, only the rotation of the space-time contributes and the angular velocity of the particle is the angular velocity $\omega_{WH}$ of the WH. So, for $\omega_{WH}$ we get 

\begin{equation}
\omega_{WH}=\frac{a}{L^3}\frac{f^2x(x^2+y^2)}{\frac{a^2}{L^4}f^2x^2(y^2-1)+(x^2+1)(x^2+y^2)^2}.
\label{rotWH}
\end{equation}

\begin{figure}[htp]
	\centering
		\includegraphics[scale=0.55]{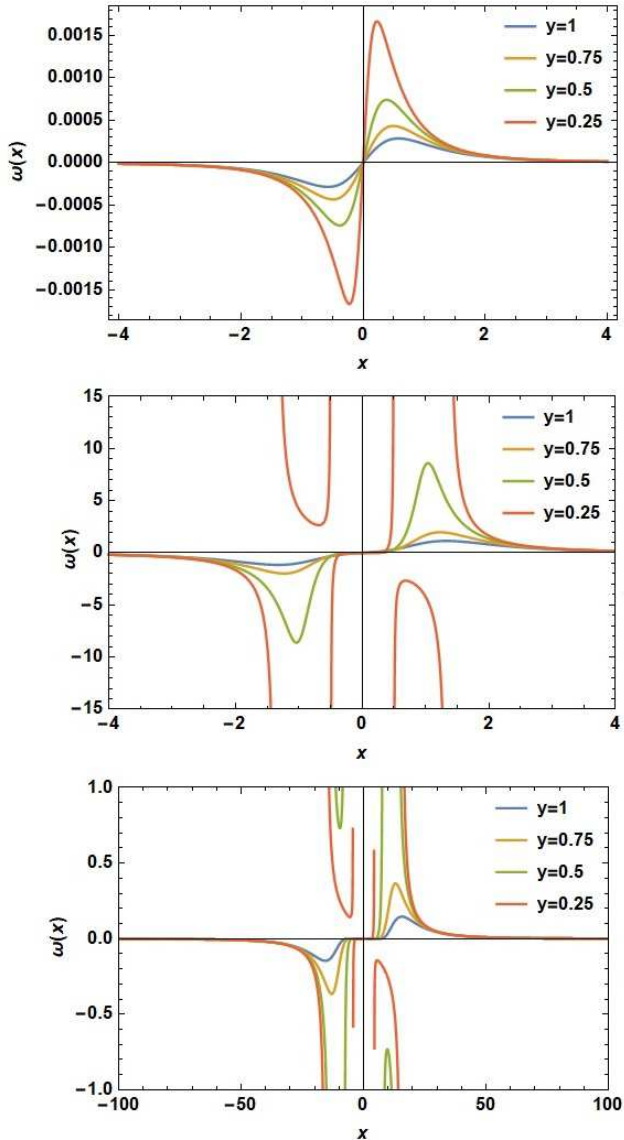} 
		\caption{Angular velocity of the WH as a function of $x$ for different values of $y$. In the top panel a case where $a\ll L^3$ with $a=0.11$, $k_1=0.1$ and $L=5$. In the middle panel the $a\sim L$ case with $a=0.11$, $k_1=0.1$ and $L=0.21$. In the bottom panel $L\ll a$ with $a=11$, $k_1=10$ and $L=0.21$. The function will grow to infinity as it approaches the ring singularity ($x=0$, $y=0$) for any case.}
	\label{fig:omega}
\end{figure}

Special attention deserves the denominator in (\ref{rotWH}). In a general case there exist non-trivial real values of $x$ and $y$ where $\frac{a^2}{L^4}f^2x_0^2(y_0^2-1)+(x_0^2+1)(x_0^2+y_0^2)^2=0$, which consequently makes the angular velocity diverge, this is not desirable. However, if we examine the case $a\ll L^3$ then the second term of this equation will dominate over the first one, that is, $\left|\frac{a^2}{L^4}f^2x^2(y^2-1)\right|<\left|(x^2+1)(x^2+y^2)^2\right|$ everywhere except near the ring singularity. Therefore, in this limit one obtains a well behaved function of $\omega_{WH}$ that also satisfies $\omega_{WH}\ll 1$ as it is shown in Fig.(\ref{fig:omega}). For the rest of this paper we shall refer to $a\ll L^3$ as the slowly rotating limit.

We may finally add to this regard that, in the rest of the cases besides the slowly rotating limit, i.e: $L\ll a$ and $a\sim L$, there is no general dominance of one term over the other in the denominator of (\ref{rotWH}). As a consequence, at some region in space-time $x=x_0$, $y=y_0$, these terms may cancel each other out leading to infinite angular velocities. It can be argued then, that if a wormhole of this type were to be formed it would most likely be characterized by slow rotations, since the other cases lead to unphysical scenarios. Then again, a WH by itself seems to be a collection of unphysical behaviors.

Considering the slowly rotating limit to first order in $a\ll L^3$ and assuming $a\sim k_1$, we can approximate the metric parameter (\ref{omegaf}) as

\begin{equation}
f\approx 1+\left(\frac{a^2-k^2_1}{2k_1}\right)\frac{y}{L^2(x^2+y^2)}.
\label{faprox}
\end{equation}

\section{III. Region of Validity of the Slowly Rotating Limit}\label{sec:val}

The first order term in approximation (\ref{faprox}), as well as the higher order terms, depend on the coordinates $x$ and $y$. Because of this, it is important to discuss in which regions of the space-time this approximation describes adequately the original function $f$ as given by expression (\ref{omegaf}).

To consider valid the series expansion for $f$ the first step would consist on checking for convergence, for this purpose Cauchy's ratio test can be proven useful. However, given the functional form of $f$ and as we are interested in a first order approximation, a more helpful criteria can be considered to be $\left|a_2/a_1\right|<1$, where $a_n$ is the n-th order term of the approximation. This is basically Cauchy's criteria with $n=1$ and will determine whether the first order term suffices to approximate $f$. For this case at least, this condition is more restrictive than the general ratio test $\lim_{n\rightarrow\infty}\left|a_{n+1}/a_n\right|<1$ so convergence is guaranteed.

The first and second order terms obtained by expanding $f$ are

\begin{eqnarray}
a_1&=&\frac{a^2-k_1^2}{a^2+k_1^2}\lambda, \nonumber\\
a_2&=&\frac{a^4-6a^2k_1^2+k_1^4}{2(a^2+k_1^2)^2}\lambda^2.
\label{a1a2}
\end{eqnarray}

So the above criteria becomes 

\begin{equation}
\left|\frac{2cy}{x^2+y^2}\right|<1,
\label{crit}
\end{equation}

with $c=(a^4-6a^2k_1^2+k_1^4)/8L^2(a^2-k_1^2)k_1$ for $a\neq k_1$.

\begin{figure}[htp]
	\centering
		\includegraphics[scale=0.4]{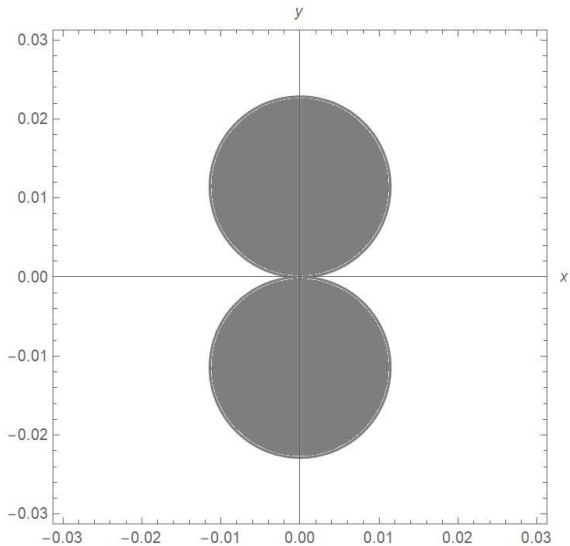}
		\caption{Validity of approximation (\ref{faprox}) in the x-y plane with $L=5$, $a=0.11$ and $k_1=0.1$, so $c=$. The gray region indicates the values of $x$ and $y$ in which the second order term of the approximation is greater than the first order term, rendering the first order approximation inaccurate.}
			\label{fig:cc}
\end{figure}

Inequality (\ref{crit}) could be rearranged to the more familiar form $x^2+(y\pm c)^2>c^2$ for $x,y\neq 0$. For the ring singularity, and very close to it, the numerator in (\ref{crit}) dominates over the denominator and hence the inequality is not satisfied. Thus, by means of the defined criteria, the domain of validity of the slowly rotating limit is described in the x-y plane by the region outside of two circles with their centers located at $(0,\pm c)$ and radius $c$. See Fig.(\ref{fig:cc}). Notice that as $k_1\rightarrow a$, $c\rightarrow \infty$ since the first order term vanishes. The particular case $a=k_1$ would need a different analysis. 

\begin{figure}[htp]
	\centering
		\includegraphics[scale=0.6]{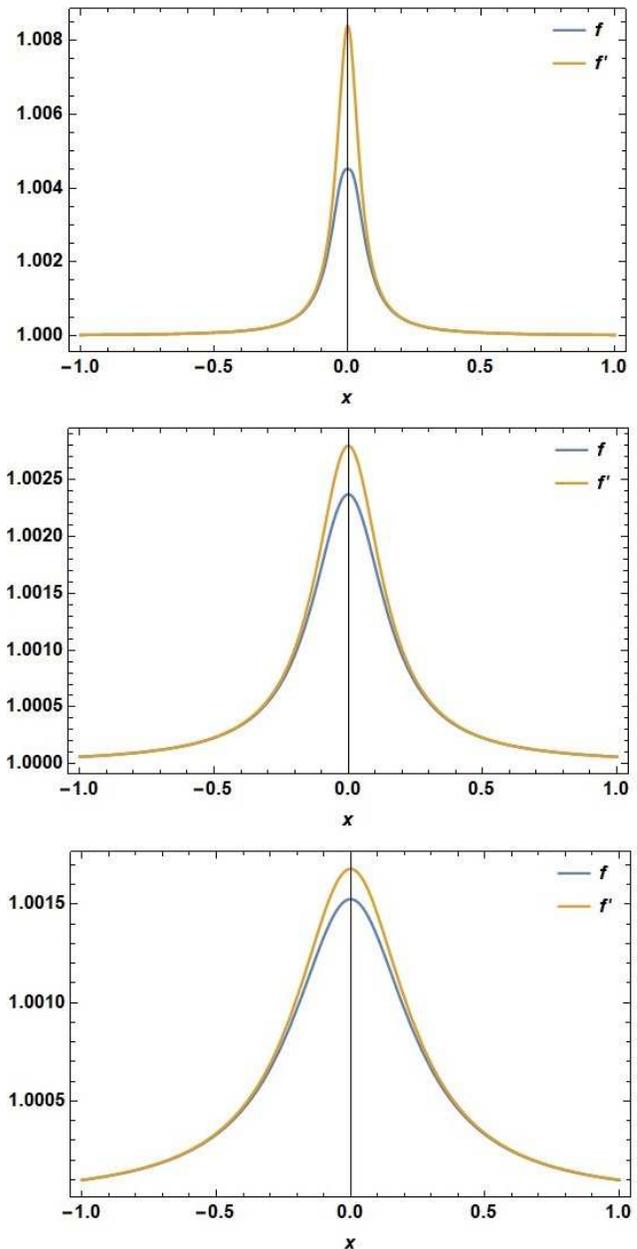} 
		\caption{ The expressions $f$ and $f'$ as a function of $x$ with fixed values of $y_0$. In this case $L=5$, $a=0.11$ and $k_1=0.1$. The values of $y_0$ for the top, middle and bottom panel are $y_0=0.05$, $y_0=0.15$ and $y_0=0.25$, respectively.}
			\label{fig:aprox}
\end{figure}

We also show in Fig.(\ref{fig:aprox}) a comparison between the behavior of the original function $f$ and the approximation $f'$ made in the slowly rotating limit. Naturally, as we get closer to the ring singularity the approximation begins to disagree with respect to the original parameter. Nevertheless, it can be seen that the valid domain of the approximation is close enough to the singularity to correctly describe the nearby region.

\section{IV. The Hamiltonian}\label{sec:ham}
From now on we work with the Hamiltonian of the geodesics. From (\ref{ds2xy}) it follows:

\begin{eqnarray}
2H=&-&\frac{p^2_t}{f}+\frac{f}{L^2(x^2+y^2)}\left((x^2+1)p^2_x+(1-y^2)p^2_y\right) \nonumber\\
&+&\frac{f(p_\phi+\Omega p_t)^2}{L^2(x^2+1)(1-y^2)}.
\label{hamiltoniano}
\end{eqnarray}

The Hamiltonian (\ref{hamiltoniano}) is itself a constant of motion: $2H=\kappa$, with $\kappa=0$ for lightlike geodesics and $\kappa=-1$ for timelike geodesics.

In the metric (\ref{ds2xy}) $\frac{\partial}{\partial t}$ and $\frac{\partial}{\partial\phi}$ are Killing vectors, so $p_t=-\mathcal{E}$ and $p_\phi=\mathcal{L}$ are constants of motion. The conserved quantities $\mathcal{E}$ and $\mathcal{L}$ are associated with the energy and angular momentum of the particle falling into the WH, respectively.

Dividing by $f$ and applying the slowly rotating limit to the Hamiltonian (\ref{hamiltoniano}), yields

\begin{eqnarray}
\frac{\kappa}{f}=&-&\frac{\mathcal{E}^2}{f^2}+\frac{(x^2+1)p^2_x+(1-y^2)p^2_y}{L^2(x^2+y^2)} \label{hamlim}\\ 
&+&\frac{\mathcal{L}^2}{L^2(x^2+1)(1-y^2)}+\frac{2a\mathcal{L}\mathcal{E}x}{L^3(x^2+1)(x^2+y^2)},
\nonumber
\end{eqnarray}

where now the approximation (\ref{faprox}) is used as $f$. By making use of the Hamilton-Jacobi theory, and after some algebraic work, it can be shown that (\ref{hamlim}) can be separated into two equations for variables $x$ and $y$. As a result, a fourth conserved quantity $\mathcal{K}$ is found:

\begin{eqnarray}
\mathcal{K}=&&-L^2(\mathcal{E}^2+\kappa)x^2+(x^2+1)p^2_x \nonumber\\
&&-\frac{\mathcal{L}^2}{x^2+1}+\frac{2a\mathcal{L}\mathcal{E}x}{L(x^2+1)} \nonumber\\
=&&L^2(\mathcal{E}^2+\kappa)y^2-(1-y^2)p^2_y \nonumber\\
&&+\left(\frac{\kappa}{2}+\mathcal{E}^2\right)\left(\frac{k^2_1-a^2}{k_1}\right)y-\frac{\mathcal{L}^2}{1-y^2}.
\label{ka}
\end{eqnarray}

The fact that there exists a fourth constant of motion is a hint that there should be a hidden symmetry in this space-time once the particular slowly rotating limit has been applied. In this case, the hidden symmetry takes the form of a second-rank Killing tensor $\mathcal{K}^{\mu \nu}$ \cite{killingt}. From (\ref{hamlim}), it can be seen that the inverse metric in the slowly rotating limit (SRL) $g_{SRL}^{\mu \nu}$ can be separated in the form

\begin{equation}
g_{SRL}^{\mu \nu}=\frac{\mathcal{X}^{\mu \nu}(x)+\mathcal{Y}^{\mu \nu}(y)}{h_1(x)+h_2(y)},
\label{ginv}
\end{equation}

with $h_1(x)=L^2x^2$ and $h_2(y)=L^2y^2+(k_1^2-a^2)y/2k_1$. Also 

\begin{eqnarray}
\mathcal{X}^{\mu \nu}&=&
\begin{bmatrix}
	-L^2x^2 & 0 & 0 & -axL/\Delta_1 \\
	0 & \Delta_1/L^2 & 0 & 0 \\
	0 & 0 & 0 & 0 \\
	-axL/\Delta_1 & 0 & 0 & -L^2/\Delta_1 \\ 	
\end{bmatrix}, \\
\label{yuv} 
\mathcal{Y}^{\mu \nu}&=&diag\left[-L^2y^2+\frac{(a^2-k_1^2)}{k_1}y,0,1-y^2,\frac{1}{1-y^2}\right].
\nonumber
\end{eqnarray}

Because $\mathcal{X}^{y\mu}=\mathcal{X}^{\mu y}=\mathcal{Y}^{x\mu}=\mathcal{Y}^{\mu x}=0$ the Hamiltonian is separable in the slowly rotating limit. Hence, the metric admits a Killing tensor given by

\begin{equation}
\mathcal{K}^{\mu \nu}=-h_1g_{SRL}^{\mu \nu}+\mathcal{X}^{\mu \nu}=h_2g_{SRL}^{\mu \nu}-\mathcal{Y}^{\mu \nu}
\label{killing}
\end{equation}

Using now the Killing tensor (\ref{killing}) the fourth constant of motion can be found by $\mathcal{K}=\mathcal{K}^{\mu \nu}p_\mu p_\nu$. One can easily verify that this calculation yields the same result for $\mathcal{K}$ as previously expressed in (\ref{ka}).

The momenta $p_x$ and $p_y$ are related to the velocities $\dot x=dx/d\tau$ and $\dot y=dy/d\tau$, where $\tau$ is an affine parameter, through

\begin{equation}
p_x=\frac{L^2(x^2+y^2)}{f(x^2+1)}\dot{x}, \hspace{0.5cm} p_y=\frac{L^2(x^2+y^2)}{f(1-y^2)}\dot{y}. 
\label{pxpy}
\end{equation}

Then, using (\ref{pxpy}) and (\ref{ka}) we have

\begin{equation}
(\Delta/f)^2\dot{x}^2=X(x), \hspace{0.5cm} (\Delta/f)^2\dot{y}^2 = Y(y),
\label{xysep}
\end{equation}

where

\begin{eqnarray}
X(x)=&&\Delta_1((\kappa+\mathcal{E}^2)x^2+\mathcal{K}/L^2) \nonumber\\
&&-2a\mathcal{EL}x/L+\mathcal{L}^2,
\label{X}\\
Y(y)=&&(1-y^2)\left[(\kappa+\mathcal{E}^2)L^2y^2-\mathcal{K} \right. \nonumber\\
&&+(\kappa/2+\mathcal{E}^2)(k^2_1-a^2)y/k_1]-\mathcal{L}^2.
\label{Y} 
\end{eqnarray}

With the help of (\ref{X}) and (\ref{Y}) a relationship between $x$ and $y$ can be found which no longer depends on the affine parameter $\tau$: 

\begin{equation}
\frac{dy}{dx}=\sqrt{\frac{Y(y)}{X(x)}}.
\label{YX}
\end{equation}

This differential equation can be solved numerically in order to describe the motion of the freely falling test particle. Nevertheless, we shall take a different approach and analyze the fourth degree polynomials (\ref{X}) and (\ref{Y}). In particular, the nature of their roots, which will give us information about the accessible regions in the WH to our traveler. 

\section{V. The Polynomials $X(x)$ and $Y(y)$}\label{sec:polXY}
We now analyze the fourth degree polynomials (\ref{X}) and (\ref{Y}). From (\ref{xysep}) it can be seen that in order to obtain real velocities, we must have for a pair of coordinates $x=x_0$ and $y=y_0$ that $X(x_0)>0$ and $Y(y_0)>0$. Otherwise, this would result on a non-physical case and the particle could never reach said coordinates in the space-time. This is the motivation behind the study of the nature of the roots of (\ref{X}) and (\ref{Y}). 

In a fourth degree polynomial the nature of its roots is determined by the discriminants $\Delta_x$, $P_1$ and $P_2$, which of course depend on the coefficients of the polynomial \cite{quartic}. In the case of (\ref{X}), these discriminants are given by:

\begin{eqnarray}
\Delta_x=&16&A(16A^2B^3-8AB^2D^2+36ABDC^2-27AC^4 \nonumber\\
&+&BD^4-C^2D^3), \nonumber\\
P_1=&8A&D, \hspace{0.6cm} P_2=16A^2(4AB-D^2),
\label{dis}
\end{eqnarray}

with $A=L^2(\mathcal{E}^2+\kappa)$, $B=\mathcal{K}+\mathcal{L}^2$, $C=a\mathcal{L}\mathcal{E}/L$ and $D=A+\mathcal{K}$.

As it will be later explained, we are interested in the scenario where $X(x)=0$ has two pairs of complex conjugate roots. For this to be the case, the conditions on the discriminants (\ref{dis}) are as follows: $\Delta_x>0$ and either $P_1>0$ or $P_2>0$.

In order to cross the throat of the WH it is necessary that $X(0)>0$, that is $B>0$. Furthermore, if the traveler were to freely move through the upper universe ($x>0$) and the lower universe ($x<0$), it would be needed that $X(x)>0$ for all $x\in \mathbb{R}$. This is accomplished by demanding that $B>0$ and that (\ref{X}) has two pairs of complex conjugate roots. 

So far the only condition for the quantities appearing in (\ref{dis}) is $\kappa=0,-1$ and $B>0$. Given this last condition notice that a term by term sign examination of $\Delta_x$ reveals that if $A<0$ and $D<0$, then $\Delta_x<0$. Performing a similar analysis, one can find that if $P_2>0$, it has to be that $A>0$ and additionally $4AB>D^2$. Also, with $A>0$, then $D>0$ so that $P_1>0$.

With this in mind, it can be stated that if $X(x)>0$ for all $x\in \mathbb{R}$, then the following conditions should be met:
\begin{enumerate}
 \item  $A>0$ (Note that this is trivially satisfied for null geodesics).
 \item  Either $D>0$ or $4AB>D^2$.
\end{enumerate}

Now, for the polynomial $Y(y)$ its discriminants are somewhat more complicated than for $X(x)$:

\begin{eqnarray}
P_1&=&-8AD-3E^2, \nonumber\\
P_2&=&16A^2(4AB-D^2)-E^2(3E^2+16A\mathcal{K}),
\label{disy}
\end{eqnarray}

with $E=(\kappa/2+\mathcal{E}^2)(k^2_1-a^2)/k_1$. It is important to mention that here we omitted a discriminant $\Delta_y$, mainly due to it being a large expression but also because independently of its sign, (\ref{Y}) can have real roots, which is what we desire for this polynomial. 

In the ring singularity the polynomials take the values $Y(0)=-X(0)=-B$. This hints that the traveler would not be able to touch the ring singularity at $x=y=0$. One should be careful though, since at exactly this region the slowly rotating approximation is not valid. However, this gives information on the behavior of the polynomials which would later allow us to reach said conclusion. Not to mention that the separate evaluation holds, i.e. $Y(0)=-B$ with some $x$ where the slowly rotating approximation is valid, and viceversa. To travel through the WH in a plane $y=y_0$, there should be some $y_0$ in which $Y(y_0)>0$. For that matter, we search for cases where (\ref{Y}) has real roots.

\begin{figure}[htp]
	\centering
		\includegraphics[scale=0.45]{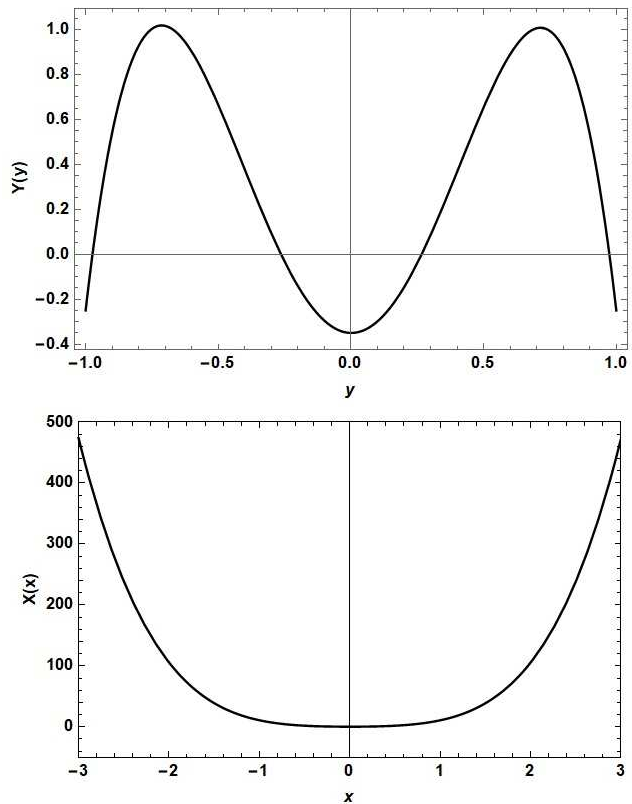} 
		\caption{The polynomials $X(x)$ and $Y(y)$ for timelike geodesics with $\mathcal{L}=0.5$, $\mathcal{E}=1.1$, $k_1=0.1$, $a=0.11$, $L=5$ and $\mathcal{K}=0.1$. The roots of $Y(y)$ are at: $y_1=-0.263578$,  $y_2=0.266584$, $y_3=-0.974137$ and  $y_4=0.973971$. }
	\label{fig:XY}
\end{figure}

The conditions on the discriminants (\ref{disy}) for four real roots in $Y(y)=0$ are: $P_1<0$ and $P_2<0$. So, if $X(x)>0$ for all $x\in \mathbb{R}$ then $A>0$, and it is sufficient to impose that: $D>0$, $D^2>4AB$ and $\mathcal{K}>0$ in order to fulfill said conditions on (\ref{disy}). See Fig.(\ref{fig:XY}).

\begin{figure}[htp]
	\centering
		\includegraphics[scale=0.45]{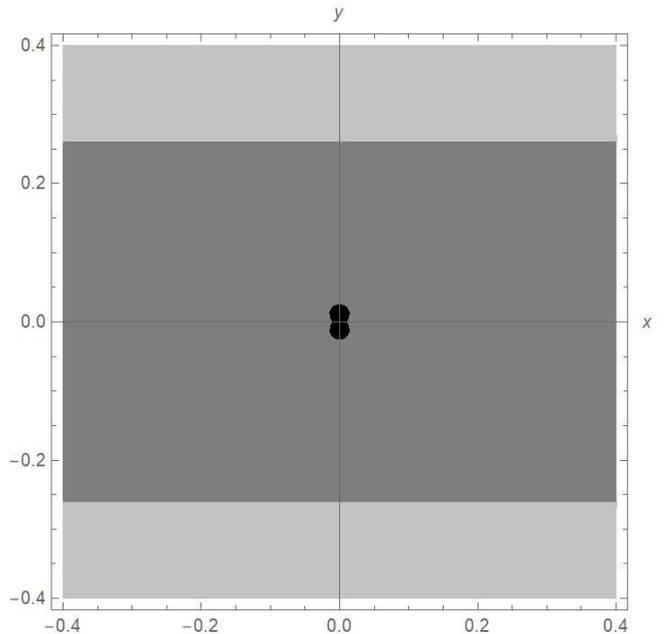} 
		\caption{Main regions of interest in the x-y plane close to the ring singularity. In the light gray region $Y(y)>0$ and so the traveler is freely allowed to move, while in the dark gray region $Y(y)<0$ and repulsive effects emerge. In the black region the slowly rotating limit is no longer valid. In this example $\mathcal{L}=0.5$, $\mathcal{E}=1.1$, $k_1=0.1$, $a=0.11$, $L=5$ and $\mathcal{K}=0.1$. }
	\label{fig:reg}
\end{figure}

At this point we may discuss the validity of our results. Even though the slowly rotating approximation breaks inside and very near the ring singularity, its valid domain is close enough from it to observe its repulsive effects, namely the negative behavior of the polynomial $Y(y)$. We conclude from this that the ring singularity is surrounded by a repulsive potential which is responsible of "censoring" it in this particular WH space-time. See Fig.(\ref{fig:reg}).

Thus we have proved that, in the slowly rotating limit, a traveler following geodesic motion in this WH could never be able to touch the singularity of the space-time. Nevertheless, satisfying the conditions on the conserved quantities studied above, it would be possible for him to freely travel back and forth both asymptotically flat universes in a plane $y=y_0$ with $Y(y_0)>0$. 

Of course, the behavior of the curves (\ref{X}) and (\ref{Y}) shown in Fig.(\ref{fig:XY}) is not unique and will change depending on the four constants of motion through the discriminants (\ref{dis}) and (\ref{disy}). However, given the form of the coefficients of these polynomials is not possible to find curves where: $X(x)>0$ for all $x\in \mathbb{R}$ and $Y(y)>0$ for all $y\in \left[-1,1\right]$.

\section{VI. Negative Mass Kerr Black Hole Comparison}   

\begin{figure}[htp]
	\centering
		\includegraphics[scale=0.45]{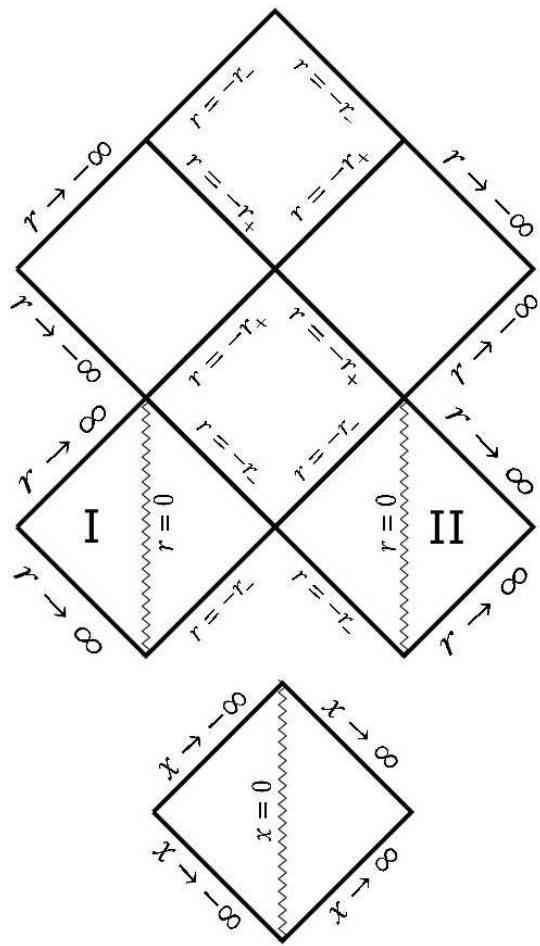} 
		\caption{Penrose diagrams for the negative mass Kerr black hole (top) and the Kerr-like phantom WH (bottom). The outer and inner horizons of the black hole are denoted by $r_+$ and $r_-$, respectively. Regions labeled as I and II share the same structure when compared to either separated universes in the WH.}
	\label{fig:diagrama}
\end{figure}

As it was mentioned earlier this wormhole possess a negative mass given by $M=-l_1$, which makes it bear certain resemblance between it and a negative mass Kerr black hole. Both space-times consist of a ring singularity and are distinguished by its lack of event horizons. However, in the negative mass Kerr black hole, crossing the region bounded by the ring singularity leads to another universe ($r<0$) with the structure of a typical Kerr space-time. Meanwhile, in this Kerr-like WH the region bounded by the ring singularity is identified as its throat and communicates universes with the same geometrical structure. It can be said, though, that the region $r>0$ of a negative mass Kerr black hole is similar to either universe of the studied WH, as they share the same Penrose diagram (see Fig.(\ref{fig:diagrama})). This last statement does not hold when viewing both space-times as a whole and comparing them.

The Penrose diagram for this WH can be shown to be simple enough. Consider a slice of the space-time described by metric (\ref{ds2}) with $\phi$ constant and $y=y_0$ also constant, this yields

\begin{equation}
ds^2=f_0\left(-dt^2+\frac{L^2(x^2+y_0^2)}{f_0^2(x^2+1)}dx^2\right),
\label{ds2diag}
\end{equation}

where $f_0=f(x,y_0)$ as given by (\ref{omegaf}) and (\ref{omegalambda}). The slice of the space-time (\ref{ds2diag}) can be expressed in the form $ds^2=f_0(-dt^2+du^2)$ by the coordinate change

\begin{equation}
u=L^2\int{\frac{x^2+y_0^2}{f_0^2(x^2+1)}dx}.
\label{u}
\end{equation}

Some important properties of this coordinate change are:
\begin{itemize}
	\item As $x\rightarrow\pm\infty$, $u\rightarrow\pm\infty$ since the integrand in (\ref{u}) has the following limit
	\begin{equation}
	\lim_{x\rightarrow\pm\infty}\frac{x^2+y_0^2}{f_0^2(x^2+1)}=1.
	\label{lim}
	\end{equation}
	\item The function $u(x)$ is everywhere regular because its integrand is also everywhere regular. This implies that we can describe the entire space-time slice with a single patch in a Penrose diagram.
\end{itemize}

Using the common conformal transformation 

\begin{eqnarray}
\psi&=&\arctan (t+u)+\arctan (t-u) \nonumber\\
\xi&=&\arctan (t+u)-\arctan (t-u)
\label{psixi}
\end{eqnarray}

leaves the sliced metric as $ds^2=F(\psi,\xi)(-d\psi^2+d\xi^2)$, where $F(\psi,\xi)=f_0\sec^2((\psi+\xi)/2)\sec^2((\psi-\xi)/2)/4$ is the conformal factor. With the transformation (\ref{psixi}), the Penrose diagram is drawn by identifying the regions $x\rightarrow\pm\infty$ as straight lines with unity slope. While $x=0$ will be represented implicitly by $u(0)=\tan(\psi+\xi)-\tan(\psi-\xi)$. We show this Penrose diagram, along with the one of a negative mass Kerr black hole, in Fig.(\ref{fig:diagrama}).

One important proposal could be made from the Penrose diagrams in Fig.(\ref{fig:diagrama}). If regions I and II in the Kerr black hole are identified with each of the universes of the WH it would be possible to speculate about its formation. Notice that if both radii of the outer and inner horizons of the black hole shrank to such a extreme so that they coincide with that of the ring singularity, the entire unlabeled regions of the Penrose diagram would disappear. We could then join I and II through the region bounded by the singularity, allowing particles to travel on causal curves from one asymptotically flat universe to another, and thus creating a wormhole. It is worthy to point out that traveling from region I to II, or vice-versa, is not possible for a physical observer in the regular case.

A strong candidate that would fulfill a task just as the one described above is accretion, namely accretion of a phantom scalar field. In \cite{acrecion} it was found that this type of accretion reduces the area of a black hole horizon and could even be considered as an evaporation process. Although, this was done for a spherically symmetrical black hole and it remains to be seen if the same results can be obtained for the axially symmetrical case. Despite this fact, we conjecture that the Kerr-like phantom WH we have studied throughout this paper can be formed as a result of a Kerr black hole accreting a phantom scalar field, hence reducing the area of its horizons and leaving a naked singularity in the space-time. This final statement has yet to be rigorously proved.

As a last remark, another interesting characteristic of the negative mass Kerr black hole is the fact that, near the ring singularity, it contains closed time-like curves when traveling in the azimuthal direction with $r$ and $\theta$ fixed, thus traveling back in time for a physical observer. Such a feature is not presented in this WH. 

\section{VII. Conclusions} \label{sec:Conclusions}
Metric (\ref{ds2}) describes the space-time of a Kerr-like phantom WH with a naked ring singularity that bounds the throat. We have shown that, through a coordinate change and applying a slowly rotating limit, the geodesic equation (\ref{hamiltoniano}) can be separated into two polynomials yielding a fourth constant of motion. Careful examination of the resulting polynomials, in particular the nature of their roots, led to the conclusion that the ring singularity is untouchable by any observer traveling in geodesic motion. This result shows that the WH contains another form of cosmic censorship, as the geodesics instead of entering into some horizon, here they are deviated to the space-time on the other side of the throat. Thus, in the same way as in a normal black hole, here the space-time singularity is causally disconnected from the rest of the universe, it is untouchable with geodesics. 
This confirms, at least for this space-time, a new concept of cosmic censorship first proposed by Roger Penrose. Despite this, it would be possible to cross back and forth both universes through the throat of the WH on a polar plane $y=y_0$. In order to do so, the inequalities on the conserved quantities here reported need to be fulfilled. Regarding the structure of the space-time, a sketch of the Penrose diagram for a slice of this WH suggests that it could be possibly formed from a Kerr black hole. \\

 \textbf{Acknowledgments.} This work was partially supported by CONACyT M\'exico under grants CB-2011 No. 166212, CB-2014-01 No. 240512, Project
No. 269652 and Fronteras Project 281;
Xiuhcoatl and Abacus clusters at Cinvestav, IPN; I0101/131/07 C-234/07 of the Instituto
Avanzado de Cosmolog\'ia (IAC) collaboration (http://www.iac.edu.mx/) 
fellowships. J.C.A. acknowledges financial support from CONACyT doctoral fellowships. \\

\end{document}